\PassOptionsToPackage{numbers,sort&compress}{natbib}
\documentclass[final,5p,times,twocolumn]{elsarticle}
\usepackage[T1]{fontenc}
\usepackage[utf8]{inputenc}

\usepackage{amsmath,amssymb,booktabs,microtype,xcolor}
\usepackage{placeins}
\usepackage{tikz,tabularx,array}
\usetikzlibrary{decorations.pathreplacing}
\usepackage[colorlinks=true,citecolor=blue!55!black,urlcolor=blue!55!black,linkcolor=blue!55!black]{hyperref}
\hypersetup{pdftitle={Efficient simulation of variational quantum circuits: representation, accuracy and CPU/GPU allocation},pdfauthor={Vojtech Novak, Radovan Recka, Ondrej Knebl, Ivan Zelinka}}
\newcommand{\LiHPauliGain}{3.45}
\newcommand{\LiHLightningGain}{8.14}

\newcommand{\MPSShallowGain}{2.73}

\newcommand{\GPUTFIMGain}{29.98}
\newcommand{\GPUHeisenbergGain}{31.31}
\newcommand{\GPUHEAGain}{21.33}
\newcommand{\GPULiHSlower}{2.35}
\newcommand{\GPUIndependentGain}{2.01}
\newcommand{\GPUPlacementGain}{3.72}
\newcommand{\NativeWorkerGain}{3.86}

\newcommand{\TFIM}{\mathrm{TFIM}}

\journal{Journal of Systems Architecture}
\begin{document}
\begin{frontmatter}
\title{Efficient simulation of variational quantum circuits: representation, accuracy and CPU/GPU allocation}
\author[cs,it4i,ku]{Vojt\v{e}ch Nov\'{a}k\corref{cor1}}
\ead{vojtech.novak.st1@vsb.cz}
\author[cs]{Radovan Re\v{c}ka}
\author[cs]{Ond\v{r}ej Knebl}
\author[cs,it4i,ku]{Ivan Zelinka}
\cortext[cor1]{Corresponding author.}
\affiliation[cs]{organization={Department of Computer Science, Faculty of Electrical Engineering and Computer Science, VSB-Technical University of Ostrava},city={Ostrava},country={Czech Republic}}
\affiliation[it4i]{organization={IT4Innovations National Supercomputing Center, VSB-Technical University of Ostrava},city={Ostrava},postcode={708 00},country={Czech Republic}}
\affiliation[ku]{organization={Department of Informatics and Statistics, Marine Research Institute, Klaipeda University},city={Klaipeda},country={Lithuania}}
\begin{abstract}
Classical circuit simulation supports quantum software development and verification and provides a baseline for assessing quantum advantage. We compare gate-based statevector simulation, direct Pauli evolution and matrix product states for molecular and spin Variational Hamiltonian Ansatze and hardware-efficient ansatze on CPU/GPU systems, varying width, depth, precision and resource allocation. In the tested implementations, direct Pauli evolution benefits complex molecular VHA, whereas GPUs accelerate wider spin and hardware efficient circuits. Matrix product state is useful when limited entanglement permits accurate compression; corrected CPU FP32 can meet the tested tolerance while costing more than FP64. Workload size, startup, worker memory and task-specific profiling can change the preferred allocation. Matched NumPy Pauli updates accelerate two-layer full LiH by \LiHPauliGain-fold; the GPU gives approximately 30-fold acceleration over 18-thread Aer for 28-qubit, four-layer TFIM, both in FP64. We provide a procedure for choosing representation, precision and worker allocation jointly, comparing the cost of complete energy evaluations under accuracy, memory and core constraints.
\end{abstract}
\end{frontmatter}

\section{Introduction}
Classical circuit simulation supports the development and verification of quantum software and provides a performance baseline for assessing quantum advantage \cite{preskill2018}. Variational quantum algorithms repeatedly evaluate parameterized circuits for energies and other observables in electronic structure, spin models and machine learning \cite{cerezo2021,kandala2017,wecker2015}. Their classical cost includes both producing a state and evaluating the requested output. We ask how to choose a simulation method and computational resources for a circuit and evaluation stream, considering execution cost, numerical accuracy and memory.

Gate-based dense simulation stores the full statevector and updates it using a sequence of gates. Qiskit Statevector and Aer share this representation but use different execution paths \cite{qiskit2024}. Expanding a long Pauli rotation into basis changes and entangling gates creates many state updates. Direct Pauli evolution \cite{pauli} applies the whole string rotation without that ladder; the NumPy \cite{harris2020array} and Lightning paths studied here use this primitive. MPS \cite{schollwock2011density} instead factorizes the state into local tensors and can be economical when intermediate entanglement permits small bonds. Bond caps and truncation cutoffs can introduce errors that must be checked for the requested observable \cite{vidal2003,sander2025}.

Hardware adds another decision. The CUDA-Q \cite{cuda_quantum_2023,cudaqsimulators} GPU adapter in this study executes a gate-expanded statevector path. Device execution can accelerate state operations, while compilation, startup and the interface also contribute to complete calls \cite{nation2025}. Resource allocation can parallelize one state or assign independent evaluations to workers. These choices have different memory and communication requirements. A low steady-state latency therefore answers only part of the question posed by a finite workload.

Qubit count sets the number of dense state amplitudes, but does not specify how much work an evaluation requires. Molecular VHA \cite{wecker2015,novak2025reliable,saoovqe} layers can contain many long Hamiltonian-derived rotations; spin circuits offer local commuting blocks; HEA uses short rotation--entangler patterns \cite{kandala2017,wecker2015,wiersema2020,illesova2025numerical}. Hamiltonian term count also changes readout cost. For MPS, useful compression additionally depends on entanglement and the tolerance \cite{cerezo2025}. Figure~\ref{fig:circuits} makes these structural differences explicit.

Repeated evaluations arise as VQE parameter candidates, shifted circuits for suitable gradient rules \cite{schuld2019}, or data samples in variational classifiers and data re-uploading models \cite{schuld2020classifiers,perezsalinas2020,illesova2025classical}. Such applications can expose independently ready calls sharing a circuit pattern. Exact statevector gradients also admit adjoint differentiation \cite{jones2020}; the allocation unit studied here is an independently requested scalar evaluation. A QML application would calibrate its own encoding and readout.

Existing benchmarks cover multicore simulators \cite{cerrudo2026}, GPU acceleration and fusion \cite{faj2023}, just-in-time kernels \cite{efthymiou2022}, and VQE/QAOA on HPC platforms \cite{asadi2024}. Vallero et al. relate circuit characteristics to statevector/tensor-network strategy \cite{vallero2026}; \textit{phase2} develops distributed native-Pauli simulation across CPUs and GPUs \cite{miller2026}. These establish structure-dependent selection and native operations. Building on them, our contribution is to demonstrate how the preferred configuration changes when representation, numerical output and parallel allocation are judged together by the cost of a finite stream of complete scalar-energy calls. The executable procedure takes a circuit/Hamiltonian, evaluation count, error tolerance and core/memory limits; checks candidate outputs and resource use; and ranks eligible measured configurations by startup plus execution cost. Task-specific profiling and broader latency-model data collection are accounted separately.

The study spans molecular/spin VHA, HEA and structural controls across widths and depths (Table~\ref{tab:study}). Matched implementation comparisons, accuracy checks and resource-allocation measurements support the procedure. Section~\ref{sec:crossmethod} translates this evidence into a practical shortlist for the tested circuit regimes, qualified by numerical references and measured configuration coverage.

\begin{table*}[t]
\centering\footnotesize\setlength{\tabcolsep}{4pt}\renewcommand{\arraystretch}{1.12}
\caption{Study overview. $n$ is qubit count; $p$ counts VHA/QAOA layers and $L$ HEA repetitions. The CPU study varies representation, FP32/FP64 precision, 1--36 threads and 1--36 workers on 36-core nodes; the GPU subset uses one/two RTX PRO 6000 Blackwell cards. The grids contain 1,087 CPU and 36 workstation configurations, each with three paired parameter seeds. Entries describe the study grid, with selected combinations rather than a Cartesian product; resource limits and reference coverage are documented in \ref{app:repro}; separate CPU validation is specified in Sec.~\ref{sec:validation}.}
\label{tab:study}
\begin{tabularx}{\textwidth}{@{}>{\raggedright\arraybackslash}p{.20\textwidth}>{\raggedright\arraybackslash}X>{\raggedright\arraybackslash}p{.235\textwidth}>{\raggedright\arraybackslash}p{.155\textwidth}@{}}
\toprule
Problem / circuit & Hamiltonian or purpose & CPU widths and layers & GPU subset\\
\midrule
\multicolumn{4}{@{}l}{\textbf{Problem-inspired VHA}}\\
Transverse-field Ising & Open-chain $-\sum ZZ-h\sum X$; $h=1$, selected $0.25,2$ & $n=8,12,\ldots,28$; $p=1,2,4,8$ & $n=20,28$; $p=4$; $n=24$ throughput$^\dagger$\\
Heisenberg & Open-chain $\sum(XX+YY+ZZ)$ & $n=8,12,\ldots,24$; $p=1,2,4,8$ & $n=24$, $p=4$\\
H2 & STO-3G/Jordan--Wigner; bond 0.74\,\AA & $n=4$; $p=1,2,4,8$ & ---\\
H4 chains, two geometries & Spacing 1.0 and 1.8\,\AA & $n=8$; $p=1,2,4,8$ & ---\\
H6 chains, two geometries & Spacing 1.0 and 1.8\,\AA & $n=12$; $p=1,2,4,8$ & ---\\
H8 chain & Spacing 1.0\,\AA & $n=16$; $p=1,2,4,8$ & ---\\
LiH, full space & Bond 1.6\,\AA; 4 electrons / 6 spatial orbitals & $n=12$; $p=1,2,4,8$ & $n=12$, $p=1,4$\\
LiH, frozen core & Same bond; 2 active electrons / 5 spatial orbitals & $n=10$; $p=1,2,4,8$ & ---\\
LiH, smaller active space & Same bond; 2 active electrons / 4 spatial orbitals & $n=8$; $p=1,2,4,8$ & ---\\
\midrule
\multicolumn{4}{@{}l}{\textbf{Hardware-efficient and controlled-entanglement circuits}}\\
RY/CNOT HEA & Local rotations + brickwork CNOTs; signed ZZ+Z readout & Main $n=10,18,24,28$; $L=1,4$, selected 16 & $n=10,24,28$, $L=4$; $n=28$, $L=16$\\
RY/RZ/CNOT HEA & Complex-amplitude counterpart; same readout & $n=10,18,24,28$; $L=1,4$ & ---\\
Controlled HEA & RY/RZ + $R_{ZZ}(\lambda)$; TFIM energy, $\lambda=0,0.05,0.25,1$ & $n=12,16,20,28$; selected $L=2,4,8,16$ & $n=24$, $L=4$; $\lambda=0.05,1$\\
\midrule
\multicolumn{4}{@{}l}{\textbf{Supplementary structural and output controls}}\\
Signed-Ising QAOA & Diagonal Ising evolution + X mixing & $n=18,24$; $p=3$ & ---\\
Product / GHZ / easy HEA & Analytic, real-amplitude and low-entanglement limits & $n=16,28,64$; shallow circuits & ---\\
Output / interface controls & Local observables, amplitude, terminal shots; reuse, batching and affinity & Selected HEA $n=8$--28 and spin/molecular cases & Local CPU baselines; independent / distributed GPU states\\
\bottomrule
\end{tabularx}
\vspace{3pt}\parbox{\textwidth}{\scriptsize All molecules use the same basis/mapping convention. The nine fixtures comprise H2, two H4, two H6, H8 and three LiH active spaces. Output controls supplement the main HEA widths with $n=8,16,20,22$. $^\dagger$The 24-qubit TFIM GPU throughput control uses FP32 only. Terminal-shot controls use 128, 1,024 and 8,192 shots.}
\end{table*}

\section{Methods}
\label{sec:methods}

\subsection{Energy evaluation and problem-inspired VHA}
The primary task is a fixed-parameter scalar energy,
\begin{equation}
 E(\boldsymbol\theta)=\langle\psi(\boldsymbol\theta)|H|\psi(\boldsymbol\theta)\rangle,
 \qquad H=\sum_j h_jP_j,
 \label{eq:energy}
\end{equation}
with Pauli strings $P_j$. We study the cost and accuracy of this evaluation, rather than optimizer trajectories or ground-state convergence. Molecular active spaces and geometries are distinct physical models; timing differences between them are not matched-unitary speedups.

For VHA, the Hamiltonian is partitioned into blocks $H_b$ whose terms commute internally. One layer applies the ordered product
\begin{equation}
 U_\ell=e^{-i\theta_{\ell B}H_B}\cdots e^{-i\theta_{\ell1}H_1},
 \qquad |\psi\rangle=U_p\cdots U_1|\phi_0\rangle.
 \label{eq:vha}
\end{equation}
If $H_b=\sum_{j\in b}h_jP_j$, its exponential is implemented by the product of rotations $R_{P_j}(2h_j\theta_{\ell b})$. Internal commutation makes that block factorization exact. The ordered layer is not an exact exponential of the sum of noncommuting blocks. Grouping and order are fixed across implementations \cite{qiskithva}. The identity coefficient contributes to energy but adds only an omitted global phase to evolution.

Spin circuits use
\begin{align}
 H_{\TFIM}&=-\sum_{j=0}^{n-2}Z_jZ_{j+1}-h\sum_{j=0}^{n-1}X_j,\\
 H_{\rm H}&=\sum_{j=0}^{n-2}(X_jX_{j+1}+Y_jY_{j+1}+Z_jZ_{j+1}).
\end{align}
TFIM starts in the plus state and uses ZZ and X blocks. Heisenberg starts in a N\'{e}el state and uses even/odd edge blocks. Their local interactions permit shorter Pauli strings and, for TFIM, a specialized commuting-block update.

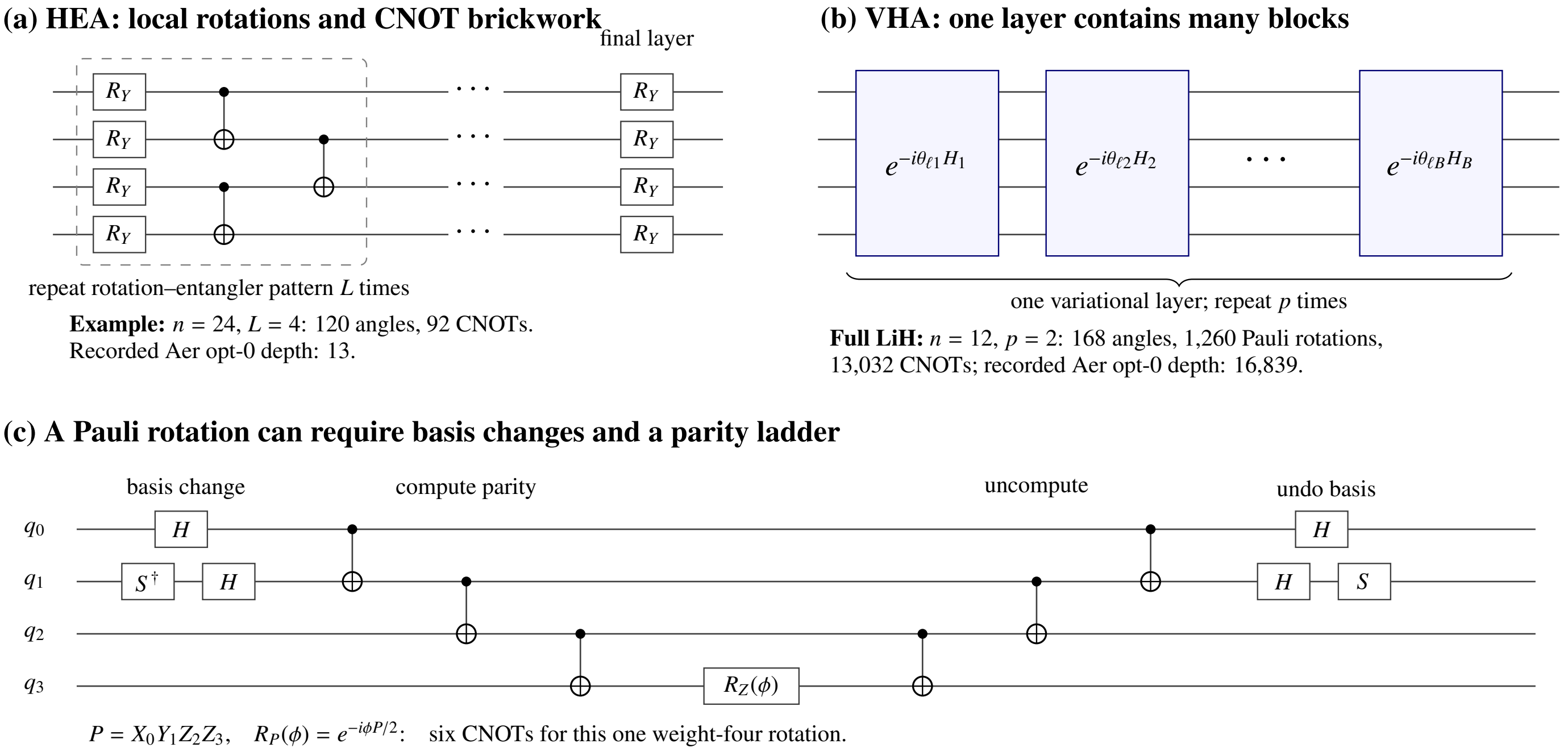
\begin{figure*}[t]
\centering
\resizebox{\textwidth}{!}{%
\begin{tikzpicture}[x=1cm,y=1cm,
  wire/.style={draw=black!75,line width=.45pt},
  gate/.style={draw=black!75,fill=white,minimum width=.55cm,minimum height=.38cm,inner sep=2pt,font=\scriptsize},
  block/.style={draw=blue!45!black,fill=blue!4,minimum height=1.95cm,minimum width=1.5cm,inner sep=2pt,font=\small},
  title/.style={anchor=west,font=\small\bfseries},
  note/.style={anchor=west,font=\scriptsize,align=left}]
\node[title] at (0,0) {(a) HEA: local rotations and CNOT brickwork};
\foreach \y in {-.75,-1.25,-1.75,-2.25} {
  \draw[wire] (.65,\y)--(7.7,\y);
  \node[gate] at (1.35,\y) {$R_Y$};
  \node[gate] at (6.9,\y) {$R_Y$};
}
\foreach \ya/\yb in {-.75/-1.25,-1.75/-2.25} {
  \draw[wire] (2.45,\ya)--(2.45,\yb);
  \fill (2.45,\ya) circle (.052);
  \draw[fill=white,line width=.55pt] (2.45,\yb) circle (.10);
  \draw (2.35,\yb)--(2.55,\yb);\draw (2.45,\yb-.10)--(2.45,\yb+.10);
}
\draw[wire] (3.5,-1.25)--(3.5,-1.75);\fill (3.5,-1.25) circle (.052);
\draw[fill=white,line width=.55pt] (3.5,-1.75) circle (.10);
\draw (3.4,-1.75)--(3.6,-1.75);\draw (3.5,-1.85)--(3.5,-1.65);
\foreach \y in {-.75,-1.25,-1.75,-2.25} {\node[fill=white,inner sep=2pt] at (5.1,\y) {$\cdots$};}
\draw[black!50,dashed,rounded corners=2pt] (.9,-.40) rectangle (3.95,-2.57);
\node[font=\scriptsize] at (2.4,-2.83) {repeat rotation--entangler pattern $L$ times};
\node[font=\scriptsize,anchor=south] at (6.9,-.44) {final layer};
\node[note] at (.7,-3.35) {\textbf{Example:} $n=24$, $L=4$: 120 angles, 92 CNOTs.\\Recorded Aer opt-0 depth: 13.};
\begin{scope}[xshift=8.6cm]
\node[title] at (0,0) {(b) VHA: one layer contains many blocks};
\foreach \y in {-.75,-1.25,-1.75,-2.25} {\draw[wire] (.10,\y)--(7.9,\y);}
\node[block] at (1.25,-1.5) {$e^{-i\theta_{\ell 1}H_1}$};
\node[block] at (3.25,-1.5) {$e^{-i\theta_{\ell 2}H_2}$};
\node[fill=white,inner sep=2pt,font=\large] at (4.85,-1.5) {$\cdots$};
\node[block] at (6.55,-1.5) {$e^{-i\theta_{\ell B}H_B}$};
\draw[decorate,decoration={brace,mirror,amplitude=4pt}] (.4,-2.65)--(7.4,-2.65);
\node[font=\scriptsize] at (3.9,-2.97) {one variational layer; repeat $p$ times};
\node[note] at (.10,-3.5) {\textbf{Full LiH:} $n=12$, $p=2$: 168 angles, 1,260 Pauli rotations,\\13,032 CNOTs; recorded Aer opt-0 depth: 16,839.};
\end{scope}
\node[title] at (0,-4.35) {(c) A Pauli rotation can require basis changes and a parity ladder};
\foreach \q/\y in {0/-5.35,1/-5.9,2/-6.45,3/-7.0} {
 \node[font=\scriptsize,anchor=east] at (.7,\y) {$q_{\q}$};
 \draw[wire] (.9,\y)--(16.25,\y);
}
\node[gate] at (2.0,-5.35) {$H$};
\node[gate] at (1.65,-5.9) {$S^\dagger$};\node[gate] at (2.5,-5.9) {$H$};
\foreach \x/\ya/\yb in {3.8/-5.35/-5.9,5.0/-5.9/-6.45,6.2/-6.45/-7.0,9.8/-6.45/-7.0,11.0/-5.9/-6.45,12.2/-5.35/-5.9} {
 \draw[wire] (\x,\ya)--(\x,\yb);\fill (\x,\ya) circle (.052);
 \draw[fill=white,line width=.55pt] (\x,\yb) circle (.10);
 \draw (\x-.10,\yb)--(\x+.10,\yb);\draw (\x,\yb-.10)--(\x,\yb+.10);
}
\node[gate,minimum width=1.0cm] at (8.0,-7.0) {$R_Z(\phi)$};
\node[gate] at (14.0,-5.35) {$H$};
\node[gate] at (13.6,-5.9) {$H$};\node[gate] at (14.45,-5.9) {$S$};
\node[font=\scriptsize] at (2.05,-4.92) {basis change};
\node[font=\scriptsize] at (5.0,-4.92) {compute parity};
\node[font=\scriptsize] at (11.0,-4.92) {uncompute};
\node[font=\scriptsize] at (14.05,-4.92) {undo basis};
\node[font=\scriptsize,anchor=west] at (.9,-7.5) {$P=X_0Y_1Z_2Z_3$,\quad $R_P(\phi)=e^{-i\phi P/2}$:\quad six CNOTs for this one weight-four rotation.};
\end{tikzpicture}
}
\caption{Circuit structure and the meaning of a variational layer. (a) A four-wire excerpt of RY/CNOT HEA, with local rotations and even/odd entanglers. (b) A VHA layer contains $B$ blocks of internally commuting Pauli terms; rotations within a block share one parameter. The numerical examples refer to the actual stated widths/layers, not matched-size circuits. Recorded Aer depths include scalar save and no hardware routing. (c) Exact elementary decomposition of one weight-four Pauli rotation. A weight-$w$ rotation uses $2(w-1)$ canonical CNOTs; applying hundreds of these rotations makes molecular VHA deep. Direct Pauli kernels implement the same rotation without these ladders.}
\label{fig:circuits}
\end{figure*}

Molecular Hamiltonians use the STO-3G basis and Jordan--Wigner mapping, with Hartree--Fock occupation reference states. The stored terms are grouped by X/Y flip mask; each block is checked for commutation and particle-number preservation. Long Pauli strings and many Hamiltonian terms make these circuits substantially deeper than the HEA examples. Figure~\ref{fig:circuits} illustrates the distinction. Full LiH has 631 Pauli terms, 84 blocks and 6,516 canonical CNOTs per VHA layer. At $p=2$, 168 block parameters control 1,260 Pauli rotations and a decomposition with 13,032 CNOTs. The 24-qubit, $L=4$ RY/CNOT HEA instead has 120 rotation parameters and 92 CNOTs. A variational-layer count is therefore not a circuit-depth measure. Detailed molecular resources are given in Table~\ref{tab:molecules}.

\subsection{HEA and structural controls}
The RY/CNOT HEA alternates a layer of independent RY rotations with nearest-neighbour even/odd CNOT brickwork, followed by a final rotation layer. Its complex counterpart adds RZ rotations. The energy operator is a fixed-seed signed ZZ+Z Hamiltonian on a ring with opposite-qubit matching, couplings $\pm1$ and local fields in $[-0.2,0.2]$. Thus HEA and VHA timings include their specified observables, rather than timing state preparation alone.

The controlled HEA replaces the CNOT entanglers with $R_{ZZ}(\lambda)$ and uses RY/RZ rotations with paired angles. Varying $\lambda$ changes entanglement while preserving the symbolic gate pattern. Its readout is TFIM energy at $h=1$. The product-state boundary at $\lambda=0$ and the deeper circuits test the useful range of MPS approximation. The QAOA, product, GHZ and local-observable controls in Table~\ref{tab:study} test structural reductions and output choices; further details are in \ref{app:controls}.

\subsection{Implementations, hardware and timing}
We compare Qiskit Statevector, Aer, PennyLane Lightning, matrix-free NumPy, and CUDA-Q. NumPy implements elementary gates, direct Pauli rotations, and a specialized TFIM block method. The direct update is
\begin{equation}
 R_P(\phi)\psi=\cos(\phi/2)\psi-i\sin(\phi/2)P\psi,
 \label{eq:pauli}
\end{equation}
using a bit permutation and phases rather than a dense operator. Lightning uses native Pauli operations; Aer uses compiled elementary-gate circuits, with molecular optimization levels 0 and 2. CUDA-Q's \texttt{nvidia} statevector target evaluates the same logical circuits after elementary-gate expansion \cite{cudaqsimulators}. Its timed output is a blocking scalar expectation, without full-state transfer. Consequently, CPU/GPU comparisons include both hardware and software representation; they do not isolate architecture alone.

The CPU platform is a dual-socket Intel Xeon Gold 6240 node with 36 physical cores and approximately 192\,GB RAM. The main physical Aer/Lightning comparisons request 18 threads; the single-state and worker studies vary the allocation explicitly. NumPy gates and BLAS use one thread. The GPU workstation has an Intel Core i9-14900KS host and two NVIDIA RTX PRO 6000 Blackwell Workstation Edition cards, each reporting 97,887\,MiB VRAM. The primary GPU execution sets GPU~0/GPU~1 power limits to 600/400\,W. The cards use x8 PCIe links and a host-bridge path. Independent GPU workers receive separate devices; a two-rank CUDA-Q/MPI mode distributes one state. Software versions and implementation details are collected in \ref{app:repro}.

Setup, first invocation, warm-up and steady evaluation are measured separately. Unless stated otherwise, timings report steady full-energy calls with reuse of the circuit, not cached objective values. Each physical instance uses three parameter-seed blocks, with the same circuit, Hamiltonian and parameter stream across implementations. We report the median across seeds and the full seed range. Benchmark speedups are geometric means of paired seed ratios; executed-stream speedups are ratios of medians across independent launches. Nested timing rounds are not independent samples. Most physical calls use two vectors, while CPU/GPU throughput panels divide a fixed global list of 36/eight vectors among workers. Local workstation CPU controls use one NumPy thread or four Aer threads.

Numerical figures and the allocation table are regenerated from recorded case-level data by the archived Python analysis, whose preparation was assisted by ChatGPT/Codex (OpenAI).

\subsection{Accuracy and resource metrics}
The primary numerical-acceptance condition is
\begin{equation}
 \Delta E=\max_{k\in\mathcal I}|E_k-E_k^{\rm ref}|\leq10^{-6},
 \label{eq:accuracy}
\end{equation}
where $\mathcal I$ contains available matched parameter indices and the reference is dense FP64. All available seed blocks must pass for an operating point to be recommended. Tolerance frontiers compare the recorded outputs with several specified error limits. The tolerance is in Hartree for molecules and model energy units for spins. It concerns simulation output, not VQE or chemical accuracy. The primary benchmark records raw SDK outputs. MPS uses bond caps 16, 64 and 256 with cutoff $10^{-10}$ \cite{aeroptions}; separate checks vary the cutoff and numerical readout as specified below. Reference coverage and independent small-state fidelity/entropy diagnostics are specified in \ref{app:accuracy}.

For $w$ workers with $t$ threads each, $wt\leq36$. Aggregate throughput $Q$ divides the fixed number of evaluations by their collective makespan. Its inverse is the completion interval per evaluation, rather than an individual request's response latency. For a finite workload on a fully reserved node, we estimate
\begin{equation}
 \widehat T_{w,t}(N)=S_{w,t}+N/Q_{w,t},\qquad
 \widehat{\rm NH}_{w,t}=\widehat T_{w,t}/3600,
 \label{eq:nodehours}
\end{equation}
where $S$ is measured startup through warm-up. We evaluate this model within each seed, then take the median. The selected sizes are multiples of the measured 36-vector CPU round. Projections assume persistent workers and constant throughput, excluding queueing and supervisory gaps. Full-node charging includes idle cores; other core budgets or charging schemes require new measurements/costs. For an assumed mean whole-node power $\bar P$ in kW, the corresponding electricity scenario is
\begin{equation}
 \widehat{\mathcal E}(N;\bar P)=\bar P\,\widehat{\rm NH}(N)\quad\text{[kWh]}.
 \label{eq:electricity}
\end{equation}
We use $\bar P=0.30$\,kW only as an illustrative constant-power scenario. Actual average power may differ between allocations; CPU-package RAPL and GPU-board telemetry are not whole-node electricity measurements.

\subsection{Finite-stream selection procedure}
\label{sec:procedure}
Inputs are a circuit/Hamiltonian/output task, $N$ independently ready parameter vectors, error limit $\epsilon$, core budget $C$, sampled-memory allowance $M$, and optional indifference fraction $\eta$. Candidates specify backend, representation, state/readout precision, MPS cap/cutoff and $w\times t$; unmeasured combinations are excluded.

Retain candidates with $wt\leq C$, $w\leq N$, sampled memory $\leq M$ and all available matched outputs within $\epsilon$. Profile startup $S$ and throughput $Q$ with objective caching disabled; rank using Eq.~\eqref{eq:nodehours}. The tested streams divide evenly among workers; the implementation uses a final-wave ceiling otherwise. Return the least-cost configuration, reference counts/records, and configurations within $(1+\eta)$ of its cost, choosing lower sampled memory within this user-defined band. Pool throughput applies only to independently ready calls.

For one workload, \emph{task-specific profiling} measures $S$ and $Q$ for a configuration shortlist. Its cost is $T_{\rm task}$, giving $T_{\rm task}+\widehat T(N)$ for profiling and one deployment; reuse amortizes this sweep.

\emph{Latency-model data collection} spans the training workloads for the optional latency fit $\widehat\ell_c=a_c+b_c\,2^{n-20}(G_c+K)$, $a_c,b_c\geq0$, uses relative-error weighted nonnegative least squares. Coefficients group hardware, backend, family, precision, threads and representation settings; $G_c$ counts updates and $K$ Hamiltonian terms. The holdout concerns fixed-precision single-state latency. The cost of collecting this broader training dataset is accounted separately from each task-specific profiling sweep.

\subsection{Targeted numerical and stream validation}
\label{sec:validation}
Targeted validation uses a 36-core Xeon Gold 6240 Barbora node and the same pinned software as the benchmark. Each of three independent launches performs task-specific profiling of NumPy and Aer at $36\times1$ and Lightning at $1\times18$, $4\times9$, $12\times3$ and $36\times1$, using 36 profiling vectors over three rounds. Profiling pools are destroyed before new pools execute 3,600 held-out vectors for NumPy, Aer and the profiling-selected Lightning allocation. Identical deployment vectors across backends and launches isolate launch variation. One-thread accuracy checks use 24 additional vectors per molecular backend or MPS setting. 

For an FP32 state $\tilde\psi$, we compare the raw SDK scalar, that scalar divided by $\nu=\langle\tilde\psi|\tilde\psi\rangle$, and
\begin{equation}
 E_{\rm norm}=\frac{\langle\tilde\psi|H|\tilde\psi\rangle_{\rm FP64}}{\nu}.
 \label{eq:normalization}
\end{equation}
An identity shift $cI$ adds $c(\nu-1)$ to the raw numerator's error and leaves normalized-energy error unchanged. The SDK scalar can additionally contain reduction error. Normalized-path timings include state export, conversion, norm computation and the specified observable reduction. MPS checks vary cutoffs $10^{-10},10^{-12},10^{-14}$ against dense FP64 on every validation vector.

\section{Results}
\subsection{Molecular VHA: native Pauli kernels outperform gate decompositions}
For two-layer, 12-qubit full LiH, Lightning's native Pauli path takes 68.76\,ms per FP64 energy call, versus 562.15\,ms with 18-thread Aer at optimization level 2: a paired \LiHLightningGain-fold gain. Aer optimization reduces its time from 760.05\,ms and its compiled depth from 16,839 to 13,602, but does not close this gap. Within single-thread NumPy, changing only the rotation implementation from elementary gates to direct Pauli updates lowers the median from 609.29 to 176.50\,ms, a paired \LiHPauliGain-fold gain. H4 improves from 58.71 to 19.18\,ms. These matched comparisons remove the basis changes and parity ladders in Fig.~\ref{fig:circuits}(c), without changing the variational circuit or energy readout.

\begin{figure*}[t]
\centering\includegraphics[width=\textwidth]{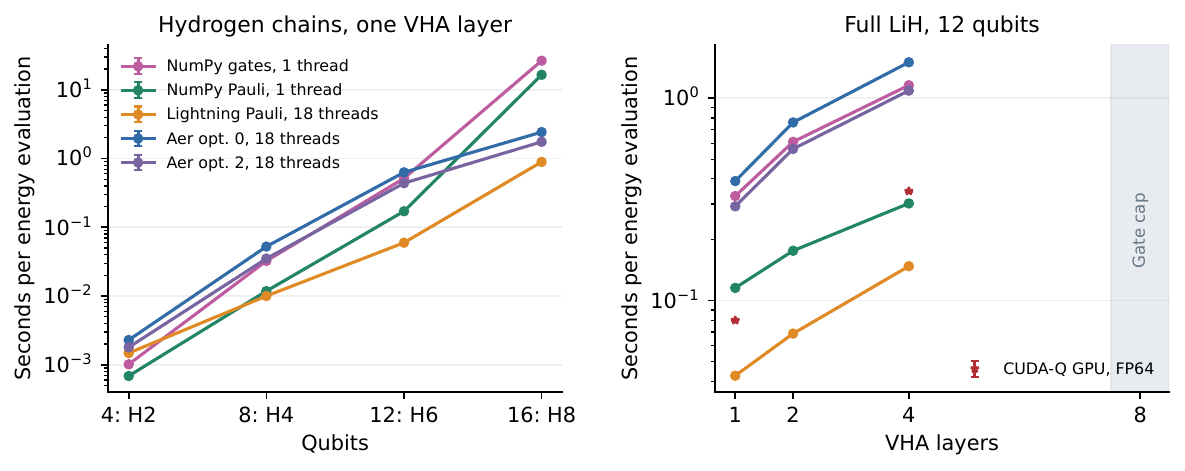}
\caption{Molecular VHA, FP64. Left: one-layer hydrogen chains at the 0.74/1.0\,\AA\ geometries in Table~\ref{tab:study}; width and Hamiltonian complexity grow together. Right: full LiH at fixed width. CPU Lightning/NumPy Pauli use direct rotations, whereas Aer and the single-GPU adapter use elementary gates. Points show three-seed medians and ranges of full-energy calls, excluding setup. The shaded eight-layer position exceeds the shared canonical-gate cap.}
\label{fig:chem}
\end{figure*}

\begin{figure*}[t]
\centering\includegraphics[width=\textwidth]{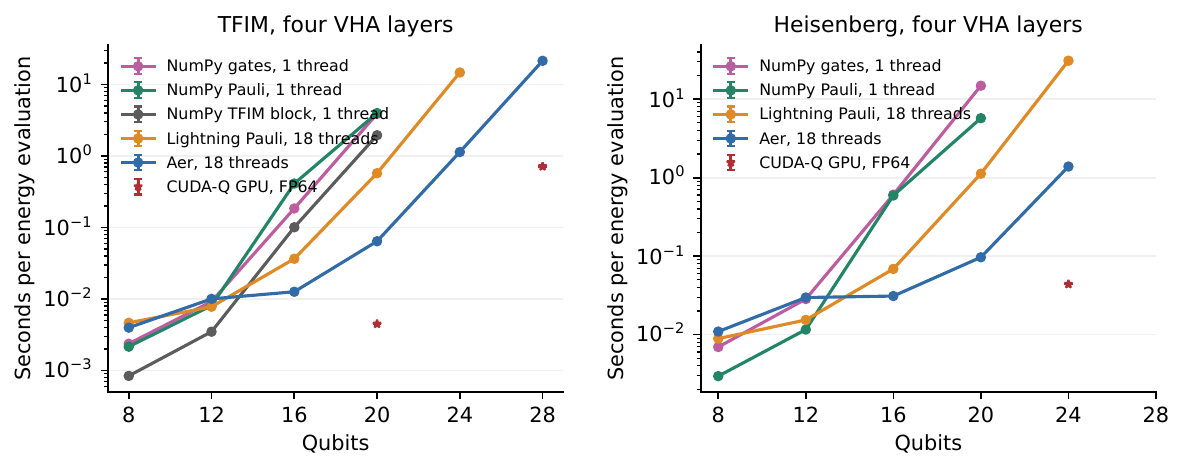}
\caption{Spin VHA, four layers, FP64. CPU curves distinguish elementary gates, direct Pauli rotations and the TFIM block method, with thread counts shown. Stars are matched single-GPU timings. Points show three-seed medians and ranges of full-energy calls, excluding setup. A point is omitted when three successful seed values are unavailable.}
\label{fig:spin}
\end{figure*}

At one layer, 16-qubit H8 contains 2,913 Hamiltonian terms and 41,600 canonical CNOTs. Its energy call takes 0.894\,s with Lightning, 1.751\,s with Aer level 2 and 16.535\,s with direct NumPy (Fig.~\ref{fig:chem}). These include evolution and full energy readout. For native NumPy LiH at four layers, the recorded mean call splits into 251.6\,ms evolution and 51.5\,ms observable evaluation (Table~\ref{tab:components}); the archive has no corresponding Lightning/GPU phase split.

The tested GPU gate path is slower than native CPU Lightning for full LiH: 79.97 versus 42.58\,ms at one layer, and 347.80 versus 147.94\,ms at four layers, giving paired CPU advantages of 1.87 and \GPULiHSlower. This compares complete implementations; a native-Pauli GPU path was not measured. Startup further penalizes short GPU workloads: the one- and four-layer paths require 6.00 and 63.42\,s through warm-up. For four layers, startup equals about 182 steady calls and remains over 10\% of the modeled total until approximately 1,642 subsequent calls.

\subsection{Spin VHA: GPU gains increase with state width}
For four-layer TFIM, the GPU lowers the 20-qubit FP64 call from 64.26\,ms with 18-thread Aer to 4.482\,ms, a paired 14.33-fold gain. At 28 qubits, 21.412\,s becomes 0.7131\,s, a \GPUTFIMGain-fold gain. The 24-qubit Heisenberg call falls from 1.386\,s to 43.88\,ms, a \GPUHeisenbergGain-fold gain (Fig.~\ref{fig:spin}).

The specialized single-thread NumPy TFIM block update takes 1.952\,s at 20 qubits, compared with 3.894\,s for elementary NumPy gates. It remains 30.41 times slower than 18-thread Aer. Thus avoiding a decomposition helps within the Python implementation but does not match the tested compiled path for this spin circuit. The workstation controls preserve the molecular/spin reversal: four-thread local Aer needs 90.62\,ms for 20-qubit TFIM versus 4.482\,ms on its GPU, whereas local direct NumPy needs only 48.47\,ms for one-layer LiH versus 79.97\,ms on the GPU.

\subsection{HEA: independent GPU workers outperform a distributed fitting state}
\begin{figure*}[t]
\centering\includegraphics[width=\textwidth]{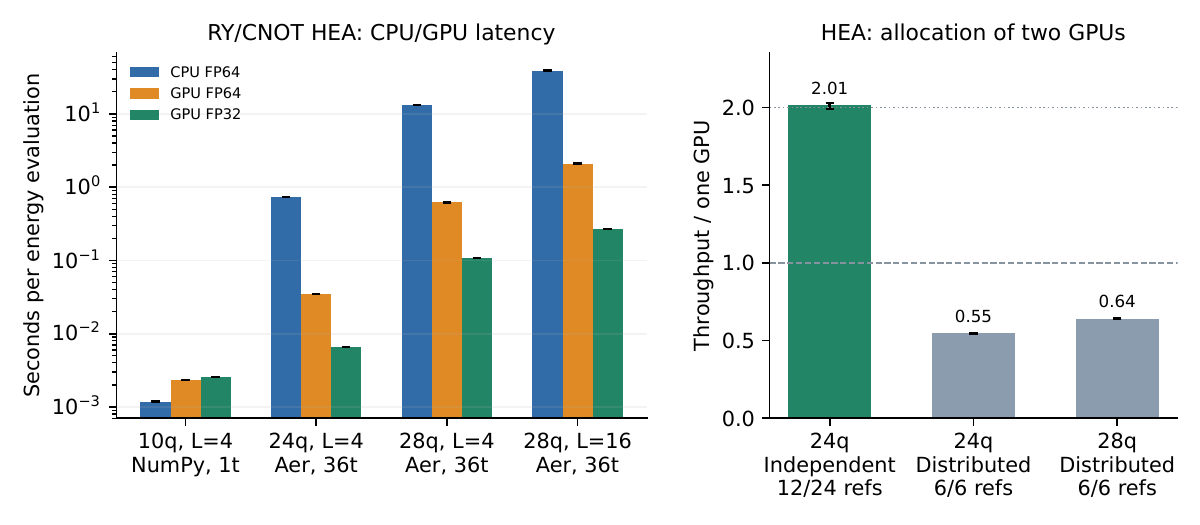}
\caption{HEA latency and GPU allocation. Left: matched full-energy calls; CPU bars use the fastest completed dense FP64 candidate in the scaling panel, named below each group. All four GPU FP32 cells pass the primary tolerance on the available compared outputs. Bars show medians and full seed ranges. Right: paired throughput ratios for two independent workers sharing eight vectors, or two GPUs distributing a state with two vectors. These use their corresponding one-GPU workloads as baselines. Lines mark equal and twice one-device throughput. Setup is excluded. Labels on the placement bars give compared/executed unique reference outputs; the independent-worker throughput point is partially checked.}
\label{fig:hea}
\end{figure*}

The ten-qubit RY/CNOT HEA takes 1.186\,ms with single-thread server NumPy, 0.409\,ms with local workstation NumPy and 2.334\,ms on the GPU in FP64. At 24 qubits and four layers, the fastest completed dense CPU configuration in the scaling panel is 36-thread Aer at 733.24\,ms; GPU FP64 takes 34.44\,ms, a paired \GPUHEAGain-fold gain. At 28 qubits, the CPU/GPU times are 13.139/0.6171\,s for four layers and 38.856/2.100\,s for 16 layers (Fig.~\ref{fig:hea}).

For the 24-qubit HEA in FP32, two independent GPU workers complete 312.48 evaluations/s, compared with 154.79 for one worker, a paired \GPUIndependentGain-fold gain. The inverse aggregate throughput is 3.200\,ms per completed evaluation; individual response latency remains approximately 6.5\,ms. Distributing one state over both cards instead takes 11.943\,ms, yielding 83.73 evaluations/s. Independent placement therefore gives \GPUPlacementGain\ times the throughput of the distributed placement on the same pair. At 28 qubits, distribution takes 167.49\,ms versus 107.45\,ms on one GPU. The compared outputs pass the primary tolerance, and both states fit one card: a 28-qubit FP32 state occupies 2\,GiB before workspace. These results concern fitting states on the tested PCIe topology.

\begin{table*}[t]
\centering\footnotesize\setlength{\tabcolsep}{3.5pt}
\caption{Projected finite-workload choices within the 36-vector FP64 worker catalogue; executed LiH results including Lightning are in Table~\ref{tab:executed}. LiH uses 12 qubits/$p=2$, TFIM 20 qubits/$p=2$ and HEA $L=4$. Each $w\times t$ minimizes projected time within its listed backend. NH and the energy scenario concern 3,600 calls; savings compare with one 36-thread Aer worker for the same task. RSS is maximum sampled process-tree memory in GiB. $^\dagger$Wh assumes a constant 300\,W whole-node power for every allocation; it is modeled electricity, not a meter reading.}
\label{tab:nodehours}\begin{tabular}{llrlrrrr}
\toprule
Workload / backend & $w\times t$, 36 & Time (s) & $w\times t$, 3,600 & NH & Saved & RSS & Wh$^\dagger$ \\
\midrule
LiH / Aer & $18\times2$ & 5.34 & $36\times1$ & 0.02510 & 96.7\% & 5.10 & 7.53 \\
LiH / NumPy Pauli & $12\times1$ & 1.18 & $36\times1$ & 0.00639 & 99.2\% & 1.36 & 1.92 \\
TFIM / Aer & $2\times18$ & 1.49 & $9\times4$ & 0.01862 & 30.4\% & 0.95 & 5.59 \\
HEA, 18q / Aer & $4\times9$ & 1.09 & $9\times4$ & 0.00832 & 46.4\% & 0.87 & 2.50 \\
HEA, 22q / Aer & $1\times36$ & 7.15 & $36\times1$ & 0.15194 & 12.1\% & 5.39 & 45.58 \\
\bottomrule
\end{tabular}

\end{table*}

\subsection{CPU allocation: executed costs and task-specific profiling}
For 3,600 full-LiH evaluations at $n=12$, $p=2$, Lightning $36\times1$ completes in 18.94\,s median across three independent Barbora deployments, versus 22.68\,s for native-Pauli NumPy and 92.07\,s for Aer (Table~\ref{tab:executed}). All three task-specific profiling sweeps select Lightning $36\times1$. Its executed makespan gives a 4.86-fold gain over Aer and 1.20-fold over NumPy, at 0.00526 versus 0.00630/0.02557 node-hours. Every deployment compares all 3,600 scalar outputs; circuit, Hamiltonian and parameter hashes match, and maximum NumPy/Lightning disagreement with Aer is $2.48\times10^{-12}$.

\begin{table}[t]
\centering\footnotesize\setlength{\tabcolsep}{3pt}
\caption{Executed two-layer LiH streams on Barbora, all $36\times1$. Times include deployment startup and result receipt, excluding task-specific profiling and worker teardown. Predicted and executed medians summarize three launches; the executed range spans all three. RSS is the maximum sampled sum of child-process RSS across launches, excluding the supervisor and potentially double-counting shared pages. Every deployment has $3,600/3,600$ matched reference outputs.}
\label{tab:executed}\begin{tabular*}{\columnwidth}{@{\extracolsep{\fill}}lrrrr@{}}
\toprule
Backend & Pred. (s) & \multicolumn{2}{c}{Executed (s)} & RSS (GiB) \\
\cmidrule(lr){3-4}
 & & Median & Range & \\
\midrule
NumPy & 23.66 & 22.68 & 21.83--23.39 & 2.78 \\
Aer & 96.06 & 92.07 & 91.99--96.38 & 6.89 \\
Lightning & 16.70 & 18.94 & 18.94--18.96 & 7.49 \\
\bottomrule
\end{tabular*}

\end{table}

Contemporaneous 36-vector forecasts differ from executed NumPy/Aer times by at most 4.8\%, but underestimate Lightning by 11.8--12.1\%. The independent 36-vector catalogue's 23.00/90.36\,s NumPy/Aer forecasts agree with executed medians within 1.9\%. The model thus selects a useful allocation without predicting every backend's makespan equally well.

Lightning's maximum sampled worker RSS is 7.49\,GiB versus NumPy's 2.78\,GiB and Aer's 6.89\,GiB. Applying a 4\,GiB sampled-memory allowance to this shortlist selects NumPy $36\times1$. Task-specific profiling intervals cost 35.66\,s median, so this sweep plus Lightning deployment costs at least 54.61\,s for one batch, exceeding direct NumPy's 22.68\,s. Reuse of task-specific profiling and the strength of the starting allocation therefore matter to this finite-workload decision.

\begin{figure*}[t]
\centering\includegraphics[width=\textwidth]{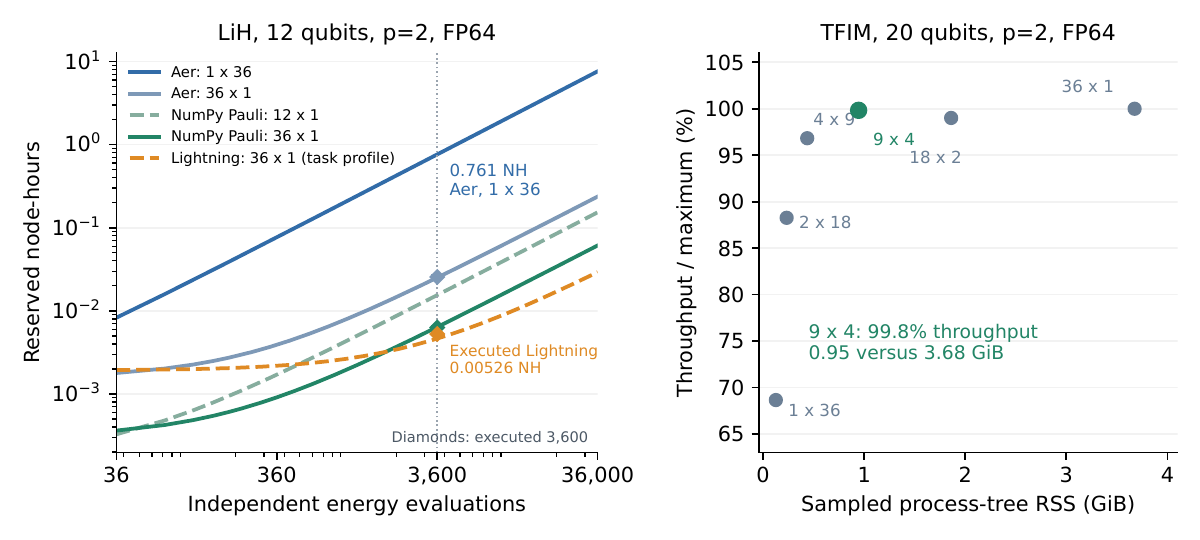}
\caption{CPU allocation on a reserved 36-core node. Left: LiH cost including startup. Solid/dashed curves and shading are medians/ranges of 36-vector projections; the Lightning curve uses the independent task-specific profiling shortlist. Diamonds show executed 3,600-vector medians and full launch ranges. Other workload sizes remain projections, excluding task-specific profiling. Right: TFIM throughput and maximum sampled process-tree RSS. Nine four-thread workers retain 99.8\% of the 36-worker throughput at 0.95 rather than 3.68\,GiB. Throughput on the right excludes startup.}
\label{fig:parallel}
\end{figure*}

The 36-vector catalogue spans more NumPy/Aer allocations (Fig.~\ref{fig:parallel}). One 36-thread Aer worker supplies 1.315 evaluations/s; $36\times1$ Aer supplies 42.459 and native-Pauli NumPy 164.249, a paired \NativeWorkerGain-fold gain over the Aer pool. For 36 calls, NumPy $12\times1$ projects 1.18\,s and the best listed Aer choice is $18\times2$ at 5.34\,s (Table~\ref{tab:nodehours}). These short-workload choices and the other table rows remain projections; the 3,600-call LiH deployments above are executed workloads.

TFIM has a flatter allocation response. At 20 qubits and two layers, $9\times4$ supplies 54.72 evaluations/s against 54.81 for $36\times1$, retaining 99.8\% with 74\% less sampled memory and shorter startup. Its median 3,600-call projection is 0.01862\,NH; the evidence supports a low-memory plateau rather than a statistically resolved unique worker count. For 22-qubit HEA, the short workload favors $1\times36$; 36 workers reduce the long-workload projection by only 12.1\%. Internal threading also differs: 24-qubit HEA falls from 19.847 to 0.734\,s in Aer from one to 36 threads. Lightning slows from 6.501\,s at nine threads to 8.038\,s at 36, while sampled memory grows from 5.45 to 18.95\,GiB.

\subsection{Latency-model selection: 33/41 fastest-candidate matches}
The affine rule trains on seeds 0/1 at $n\leq22$, depth $\leq4$, and assesses seed 2 at $n\geq24$ or depth $>4$. Table~\ref{tab:selector} compares its 33/41 fastest-candidate matches with fixed ordering and a zero-intercept fit. Fixed rules prefer an available 18-thread configuration and use deterministic fallbacks. Regret divides selected observed latency by the fastest candidate's observed latency. On the 24 binary choices, the affine fit and Aer-first rule both score 19/24; on the 17 cases with three to five candidates, they score 14/17 and 8/17, respectively.

\begin{table}[t]
\centering\footnotesize
\caption{CPU single-state selection on 41 held-out cases, with precision fixed per case. All rules use identical available configurations: 24 cases have two candidates, three have three, five have four, and nine have five. Counts include the chosen configuration and reflect eligible, successful options with fitted models. Geometric and worst regret cover all 41 cases.}
\label{tab:selector}\begin{tabular}{lrrr}
\toprule
Rule & Correct/41 & Geo. regret & Worst \\
\midrule
Affine fit & 33 & 1.10 & 3.28 \\
Zero-intercept fit & 30 & 1.22 & 5.54 \\
Aer first & 27 & 1.50 & 9.03 \\
Lightning first & 15 & 2.75 & 22.32 \\
\bottomrule
\end{tabular}

\end{table}

All eight affine errors occur at depth eight: worst regret is 3.28 for 12-qubit TFIM, selecting 18-thread Aer over specialized NumPy; Heisenberg regret is 2.74. The 38 latency models use 765 successful fitting records whose collection intervals total 1.37 node-hours. The \emph{latency-model data-collection cost} over all 972 training-domain attempts, including 20 unsuccessful executions, is 1.63 recorded node-hours, excluding allocation gaps. This broader dataset-construction expense is separate from the 35.66\,s \emph{task-specific profiling sweep} for the LiH shortlist. The test validates conditional latency selection, not accuracy or worker-scaling prediction.

\begin{figure*}[t]
\centering\includegraphics[width=.95\textwidth]{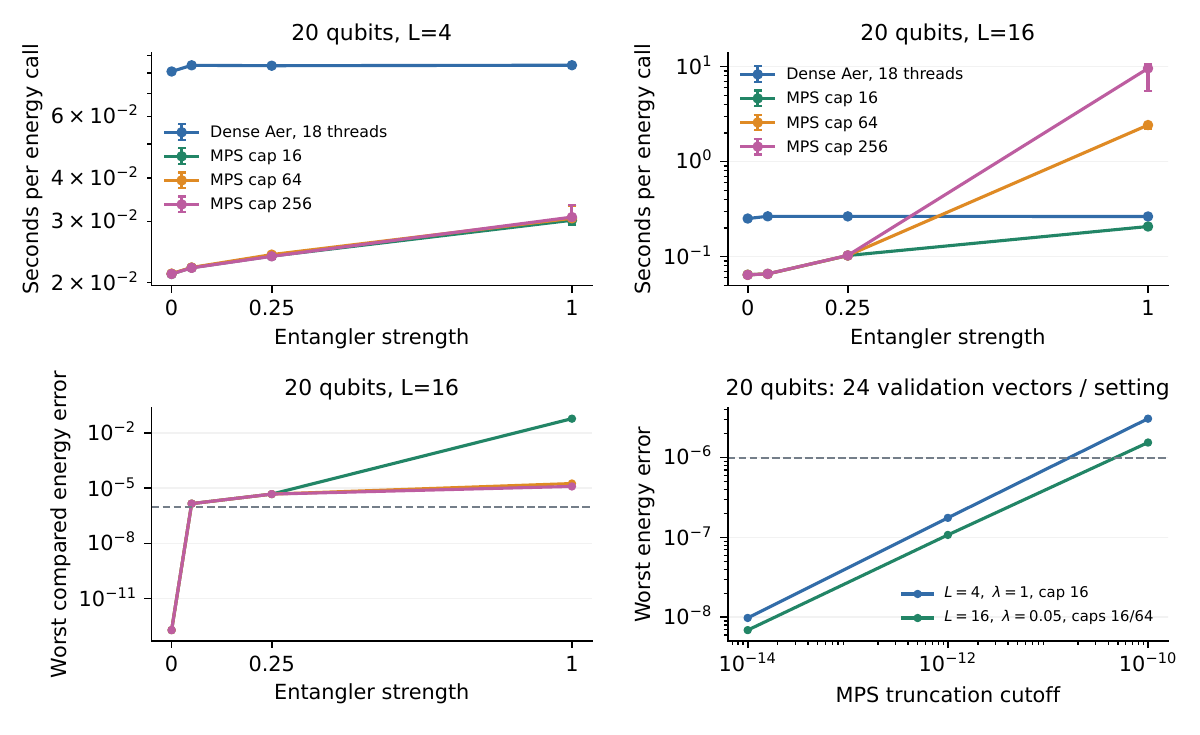}
\caption{Controlled HEA, FP64. Top and bottom left: Barbora timings and errors at cutoff $10^{-10}$, with three-seed timing medians/ranges and a $10^{-6}$ error line. Bottom right: maximum errors over 24 additional vectors per setting on Barbora as the cutoff is tightened; cap-16 and cap-64 weak-entanglement results coincide. Every validation output has a dense FP64 counterpart. The panel tests cutoff sensitivity, rather than inferring it from small-state entropy.}
\label{fig:entanglement}
\end{figure*}

\begin{figure*}[t]
\centering\includegraphics[width=\textwidth]{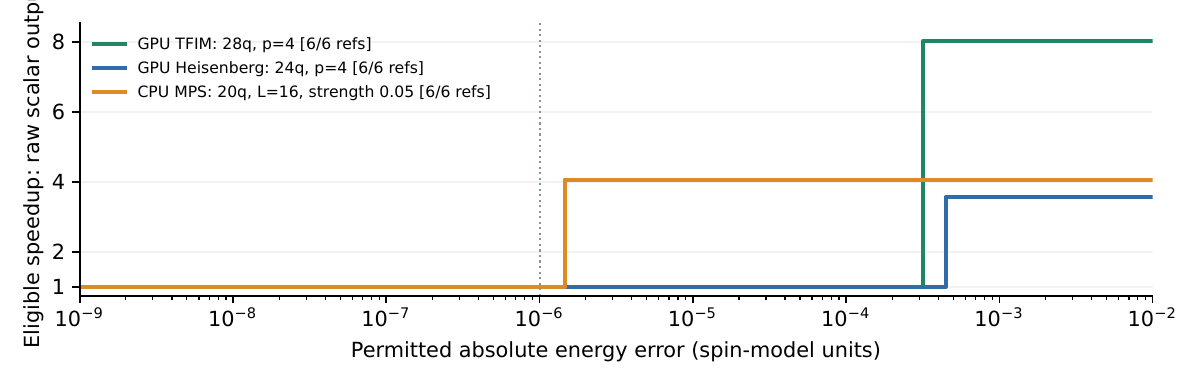}
\caption{Accuracy-dependent choices among the recorded raw-output candidates at the primary settings. GPU curves compare FP32 with FP64; the CPU curve compares fixed-cutoff MPS with dense Aer. Each legend gives compared/executed unique outputs, pooled across three seeds: $6/6$ for each displayed frontier. Gains are paired geometric ratios with setup excluded. These curves do not include the normalized CPU readout or tighter-cutoff validation, and do not certify unseen parameter vectors.}
\label{fig:precision}
\end{figure*}

\begin{table*}[t]
\centering\footnotesize\setlength{\tabcolsep}{4.5pt}
\caption{Two-layer CPU chemistry on Barbora, 24 vectors per backend. Maximum absolute error against a common dense FP64 reference. The last column is the paired geometric mean time of FP32 state evolution with normalized FP64 readout divided by the same backend's FP64 SDK time; it includes host state export and readout. A ratio above one is slower. All calls use one thread.}
\label{tab:normalization}\begin{tabular}{llrrrr}
\toprule
System & FP32 backend & Raw SDK error & SDK$/\|\psi\|^2$ error & FP64 readout$/\|\psi\|^2$ error & Time / FP64 \\
\midrule
H4 & NumPy & $3.12\times10^{-6}$ & $4.46\times10^{-7}$ & $4.46\times10^{-7}$ & 1.03 \\
H4 & Aer & $1.45\times10^{-4}$ & $1.87\times10^{-6}$ & $1.89\times10^{-6}$ & 1.05 \\
H4 & Lightning & $9.84\times10^{-6}$ & $1.01\times10^{-6}$ & $5.30\times10^{-7}$ & 1.04 \\
LiH & NumPy & $3.03\times10^{-5}$ & $2.25\times10^{-7}$ & $2.25\times10^{-7}$ & 1.03 \\
LiH & Aer & $2.35\times10^{-3}$ & $1.32\times10^{-6}$ & $1.15\times10^{-6}$ & 1.12 \\
LiH & Lightning & $6.98\times10^{-5}$ & $1.10\times10^{-5}$ & $5.64\times10^{-7}$ & 1.24 \\
\bottomrule
\end{tabular}

\end{table*}

\subsection{Controlled HEA: cutoff and validation change eligibility}
For 20 qubits, four layers and entangler strength $\lambda=1$, dense Aer takes 84.20\,ms while MPS with cap 16 takes 30.23\,ms, a paired \MPSShallowGain-fold gain. Its error on the six recorded reference outputs is $9.94\times10^{-7}$, narrowly below the primary tolerance. Caps 64 and 256 give similar times and errors (Fig.~\ref{fig:entanglement}).

At 16 layers and $\lambda=1$, cap 16 takes 0.207\,s versus 0.264\,s for dense Aer but leaves error $6.15\times10^{-2}$. Caps 64/256 cost 2.426/9.594\,s and leave errors $1.81\times10^{-5}$/$1.26\times10^{-5}$. At $\lambda=0.05$, the default cutoff leaves $1.46\times10^{-6}$ error at every cap; the recorded 4.06-fold gain is eligible at $10^{-5}$ (Fig.~\ref{fig:precision}).

On 24 additional vectors, $L=4$, $\lambda=1$, cap 16 reaches $3.08\times10^{-6}$ error at cutoff $10^{-10}$; cutoff $10^{-12}$ reduces it to $1.76\times10^{-7}$. For $L=16$, $\lambda=0.05$, errors are $1.55\times10^{-6}$/$1.08\times10^{-7}$. Observed bond is 16 throughout; cap 64 leaves matched errors unchanged. In the one-thread check, cap-16 medians at cutoffs $10^{-10}/10^{-12}$ are 68.6/71.9\,ms for four layers and 428.2/426.4\,ms for weakly entangling 16-layer circuits. Tightening cutoff therefore restores $10^{-6}$ eligibility at fixed observed bond.

\subsection{Precision: raw GPU errors and normalized CPU readout}
\label{sec:precision}
All four generic single-GPU HEA FP32 cells meet $10^{-6}$ on recorded raw outputs (Fig.~\ref{fig:hea}). For 28-qubit TFIM, FP32 lowers 0.7131 to 0.08885\,s, an 8.01-fold paired gain with $3.17\times10^{-4}$ error. Heisenberg gains 3.56-fold at $4.46\times10^{-4}$. Both pass $10^{-3}$ on six paired outputs each (Fig.~\ref{fig:precision}); normalized GPU paths are unmeasured.

LiH GPU FP32 errors reach $2.02\times10^{-3}$/$8.01\times10^{-3}$ at one/four layers, with time ratios 0.97/1.02, providing no useful precision-performance tradeoff at these settings.

CPU chemistry shows why raw-output rejection is insufficient to characterize a precision. On 24 validation vectors per backend, NumPy LiH FP32 error falls from $3.03\times10^{-5}$ to $2.25\times10^{-7}$ with normalized FP64 readout (Table~\ref{tab:normalization}). Lightning's SDK scalar divided by the same-template norm still reaches $1.10\times10^{-5}$; FP64 observable accumulation reduces this to $5.64\times10^{-7}$. Aer retains $1.15\times10^{-6}$ after that correction. The corrected paths reclassify NumPy/Lightning accuracy on these vectors. For LiH, these complete paths require approximately 3\% and 24\% more runtime than their matched FP64 SDK paths; Aer requires 12\% more runtime and still fails the tolerance. State export and readout costs prevent treating correction as a free scalar division.

\subsection{Cross-method comparison and implications for simulation selection}
\label{sec:crossmethod}
Table~\ref{tab:methodselection} translates the cross-method results into an initial shortlist. For molecular VHA with many long Pauli rotations, test native-Pauli implementations first: the matched NumPy comparison isolates the representation gain, while Lightning shows the value of its compiled implementation. For wider dense spin/HEA circuits, include GPU statevector execution alongside threaded Aer; the tested crossover depends on circuit family and implementation, so width alone cannot set a universal GPU threshold. For weakly entangling circuits, include MPS, choosing its cutoff using reference errors rather than bond cap alone.

\begin{table*}[t]
\centering\footnotesize\setlength{\tabcolsep}{4pt}\renewcommand{\arraystretch}{1.15}
\caption{Cross-method synthesis for constructing a simulation shortlist. Findings concern the implemented paths and tested circuit families; qualifications delimit their use.}
\label{tab:methodselection}\begin{tabularx}{\textwidth}{@{}>{\raggedright\arraybackslash}p{.17\textwidth}>{\raggedright\arraybackslash}X>{\raggedright\arraybackslash}X@{}}
\toprule
Simulation approach & Main finding & Important qualification\\
\midrule
NumPy direct Pauli \cite{harris2020array} & Avoiding elementary-gate ladders improves matched molecular VHA calls (Fig.~\ref{fig:chem}). & The specialized NumPy spin update still trails compiled Aer; avoiding decomposition alone does not establish the fastest implementation.\\
PennyLane Lightning \cite{asadi2024} & Strong native-Pauli molecular performance; the selected pool wins the executed LiH comparison (Fig.~\ref{fig:chem}, Table~\ref{tab:executed}). & The selected LiH pool uses more sampled RSS than NumPy. Task-specific profiling adds a separate cost.\\
Qiskit Aer \cite{qiskit2024} & Compiled dense execution performs strongly on spin/HEA cases; threading and worker pools serve different workloads (Figs.~\ref{fig:spin}, \ref{fig:parallel}). & Elementary-gate molecular execution remains slower than the tested native-Pauli competitors, even with the reported compilation optimization.\\
CUDA-Q GPU \cite{cuda_quantum_2023} & Large gains for wider spin/HEA cases; independent workers favor fitting states (Figs.~\ref{fig:spin}, \ref{fig:hea}). & The gate-expanded GPU path loses to native CPU LiH. Startup and the tested PCIe topology delimit the allocation result.\\
Aer MPS \cite{vidal2003,schollwock2011density} & Selected shallow or weakly entangling HEA permits economical approximation (Fig.~\ref{fig:entanglement}). & Bond cap, cutoff and reference-vector coverage determine eligibility; increasing the cap alone need not meet the tolerance.\\
\bottomrule
\end{tabularx}

\end{table*}

For large batches of independently ready evaluations whose states fit per worker, compare worker pools with internally parallel execution. The measured CPU/GPU placement cases favor concurrency for such workloads, whereas short streams can favor lower startup costs. Validate each candidate's complete energy-output path, including any precision correction, then rank feasible configurations for the required evaluation count and memory budget. Reuse suitable profiling where possible; a fresh task-specific sweep is useful when its expected savings justify its cost. The CPU procedure ranks measured host configurations; the GPU controls compare uses of the same device pair.

\section{Discussion}
\subsection{Representation sets the hardware workload}
Direct Pauli evolution removes basis changes and parity ladders while preserving the rotation and scalar-energy task. The matched NumPy comparisons therefore identify a representation benefit without changing hardware. Lightning supplies another native-Pauli CPU implementation with strong molecular performance. This does not make direct updates sufficient for every family: the specialized NumPy TFIM block path still trails compiled Aer. The primitive operations and their implementation both matter.

The molecular CPU/GPU reversal follows the same reasoning. Native CPU Pauli operations are compared with a GPU elementary-gate path, so the result selects a complete implementation. The large GPU gains for wider spin and HEA circuits show a different operating regime. A native-Pauli GPU comparison would be needed to separate architecture from representation in the molecular case; the available component timings do not isolate GPU or Lightning kernel costs.

\subsection{Width and layer count do not specify evaluation complexity}
At fixed precision, width determines dense statevector storage, but gate structure and observable complexity determine the work performed on that state. The circuit schematic contrasts a 12-qubit, two-layer LiH circuit with 13,032 canonical CNOTs against a 24-qubit, four-layer HEA with 92. Molecular energies also require many Pauli terms, whereas the HEA readout is comparatively sparse. These are different tasks, not matched-unitary speedups. Their resource counts explain why a smaller register can still present a costly evaluation. For MPS, intermediate bond growth adds another dimension: nominal width or depth alone does not establish compressibility.

\subsection{The workload determines the useful allocation}
Allocation should follow the application's dependence structure. Single-call latency measures how soon one requested value is available; aggregate throughput measures completed values per second when several are ready. The independent-GPU control increases throughput while individual response latency remains approximately unchanged. An optimization step that must wait for the next energy value can therefore favor a different allocation from a batch of ready parameter vectors. Internal threading helps wide Aer HEA calls, while independent CPU workers exploit ready LiH evaluations. TFIM's near-equal-throughput allocations make memory and startup useful selection criteria within the measured plateau. Distributing one state addresses a different capacity/allocation objective and is slower for the tested fitting states.

For a finite stream, compare startup plus execution time and include task-specific profiling when a fresh sweep is required. LiH illustrates the trade-off: Lightning has the shortest executed batch time, but the sweep costs more than its gain over an already-native NumPy pool for one batch, and its worker memory is higher. Repeated workloads can amortize that expense; a memory allowance can instead select NumPy. Latency-model data collection builds a broader reusable timing catalogue and remains a separate cost. Selection therefore needs the evaluation count, available independent calls and opportunities to reuse startup and profiling, in addition to per-call timing. Conditional electricity scenarios follow the same accounting (\ref{app:power}).

\subsection{Accuracy is part of the configuration}
MPS truncation and FP32 readout show how numerical settings change the eligible comparison set. The selected HEA checks recover the target accuracy by tightening the cutoff at the same observed bond dimension, while deeper strongly entangling cases remain inaccurate at the tested settings. A larger bond cap alone therefore need not remove the error floor. Raw FP32 chemistry errors also combine state evolution, norm drift and observable reduction. Normalized FP64 readout qualifies the tested NumPy/Lightning FP32 LiH paths, but requires approximately 3\% and 24\% more runtime than their FP64 paths. The relevant cost includes every operation needed to meet the tolerance.

These interpretations concern the tested circuits, implementations and available references. Three LiH deployment launches share one Barbora node and the same held-out stream. Native-Pauli GPU execution and normalized GPU readout remain unmeasured, and the canonical-CNOT cap limits the circuit grid. Fixed-parameter agreement assesses simulation output; optimizer convergence and ground-state accuracy require separate evidence.

\section{Conclusion}
The tested circuit families favor different simulation pipelines. Direct Pauli implementations benefit molecular VHA, compiled dense and GPU paths perform strongly on wider spin/HEA cases, and MPS is useful where entanglement and validated truncation permit compression. These changes in ranking support selection by circuit structure, observable and implementation, rather than by backend name or qubit count alone.

Representation, precision and parallel allocation must be considered jointly: numerical corrections change evaluation cost, and worker choices change memory and throughput. Single-circuit latency consequently does not determine the reserved cost of repeated evaluations; startup, workload size and task-specific profiling also enter the decision.

For HPC deployment, validate the complete energy-evaluation path first, then compare worker/thread allocations for the available independent calls and memory budget. Reuse task-specific profiling where suitable and charge any new sweep to the planned stream. The practical target is the total reserved cost of delivering the required accurate evaluations.

\FloatBarrier
\section*{Declaration of generative AI and AI-assisted technologies in the manuscript preparation process}
OpenAI ChatGPT/Codex was used for programming and coding: implementing and debugging the analysis code. All code, computations and results were rigorously reviewed by the authors, who take full responsibility for the content of this article.
\section*{Data and Code Availability}
The benchmark data and original study code are openly available in the Zenodo repository at \url{https://doi.org/10.5281/zenodo.23857800}.

\section*{Acknowledgements}
This project has received funding from the Research Council of Lithuania (LMTLT), agreement No.: S-ITP-25-7. This work was supported by the European Union under the REFRESH – Research Excellence For REgion Sustainability and High-tech Industries project, project number CZ.10.03.01/00/22\_003/0000048, via the Operational Programme Just Transition. This research was also supported by research grants SGS No. SP2026/063 of VSB-Technical University of Ostrava, Czech Republic. The CPU calculations used the Barbora system at IT4Innovations National Supercomputing Center under project FTA-26-86. This work was supported by the Ministry of Education, Youth and Sports of the Czech Republic through the e-INFRA CZ (ID:90254 ).

\appendix
\setcounter{table}{0}
\setcounter{figure}{0}
\section{Reproducibility and supplementary details}
\label{app:repro}
\subsection{Circuit parameters and implementation settings}
Independent angles are uniform in $[-\pi,\pi]$ with seeds $1729+1009r$, $r=0,1,2$; selected generic controls additionally use nearby trajectories and single-rotation shifts. A two-shift rule is not assumed for tied VHA block parameters. Qubit zero is the least significant statevector bit. Matched comparisons check the circuit, Hamiltonian, output task and identical parameter prefixes. Molecular fixtures omit coefficients with magnitude at most $10^{-10}$ and record an omitted-coefficient L1 bound. They were generated before timing with PennyLane's differentiable restricted Hartree--Fock implementation.

CPU software is Python 3.11.4, NumPy 2.2.6, SciPy 1.15.3, Qiskit 2.6.1, Aer 0.18.2, PennyLane 0.45.1 and Lightning 0.45.0. The workstation uses Python 3.12.3 and the same NumPy/Aer versions, with CUDA-Q's \texttt{amd64-cu13-0.16.0} build, CUDA 13.0 container environment and driver 595.99.02. Affinity and thread limits are applied before SDK import; Lightning numerical binding disables objective memoization. CPU workloads use independent single-node execution, with paired problem/seed blocks on the same node.

Most physical cells evaluate two vectors over three to eight steady rounds, targeting 0.5\,s of work within fixed caps. The common 50,000-CNOT canonical limit applies to every representation, including direct Pauli methods. H8 at $p\geq2$, H6 at $p=8$ and full LiH at $p=8$ exceed that bound. Preparation has a 120\,s deadline; most physical cells have a 300\,s total limit, selected worker cells 600\,s. A sampled process-tree memory guard uses 128\,GiB or 80\% of available memory, whichever is smaller. Such cases are censored, not plotted as measured slow times. Memory sampling may miss short peaks or double-count shared pages.

\subsection{Molecular resources}
Table~\ref{tab:molecules} details the seven resource patterns across the nine molecular fixtures. Counts refer to the canonical elementary decomposition before optimization. The compiled-depth examples in Fig.~\ref{fig:circuits} are the recorded Aer template depths, including the scalar-save instruction and without hardware routing; they are not physical device depths.
\begin{table}[t]
\centering\small
\caption{Molecular VHA resources per canonical layer. $K$ includes the identity; $B$ counts stored blocks. Each H4/H6 row represents both chain spacings.}
\label{tab:molecules}
\begin{tabular}{lrrrr}
\toprule
System & $n$ & $K$ & $B$ & CNOTs/$p$\\
\midrule
H2 & 4 & 15 & 2 & 36\\
H4, two spacings & 8 & 185 & 27 & 1,328\\
H6, two spacings & 12 & 919 & 148 & 9,972\\
H8 & 16 & 2,913 & 501 & 41,600\\
LiH full & 12 & 631 & 84 & 6,516\\
LiH frozen core & 10 & 276 & 35 & 2,362\\
LiH active 2e/4o & 8 & 105 & 13 & 472\\
\bottomrule
\end{tabular}
\end{table}

\subsection{Coverage and computational accounting}
The CPU grid records 3,261 outcomes: 2,823 executions, 318 time limits, 90 gate exclusions, 21 memory stops and nine software errors. The GPU extension records 108 outcomes per execution of its plan, comprising 84 GPU cases and 24 local CPU controls; 105 execute, and the three time limits are local NumPy controls. CPU production uses 53.93 node-hours across eight single-node allocations, including 33.76 node-hours in timed-out case intervals. The targeted validation driver records another 839.7\,s (0.233 node-hours), covering preflight, task-specific profiling, deployments and accuracy checks. These accounting totals describe the benchmark campaign rather than a simulation kernel. Large Lightning allocations trigger the memory stops; the software errors are 64-qubit easy-state MPS controls rejected by the reported 63-qubit transpilation target.

\subsection{Reference coverage and state diagnostics}
\label{app:accuracy}
Reference comparisons use common outputs rather than assuming that a self reference establishes agreement. The native-Pauli/Aer LiH worker comparison additionally matches all 36 outputs per seed and verifies their circuit, Hamiltonian and parameter hashes; all 108 agree within $2.30\times10^{-12}$. Thirty of the 33 single-GPU FP64 cells have independent CPU-backend comparisons, with maximum scalar difference $1.43\times10^{-11}$. The three controlled-HEA cells at $n=24$, $\lambda=0.05$ lack an independent FP64 counterpart; their FP32 errors are compared with the same adapter's FP64 output. The eight-output GPU throughput cells have independent references for four HEA outputs or two TFIM outputs per seed. Acceptance applies to that available subset; it does not certify unobserved vectors. A failing subset is sufficient to reject an operating point. Across the declared grids, 370 of 659 compared CPU FP32 cells and 26 of 51 GPU FP32 cells exceed tolerance; related placements are not independent circuit samples.

The largest CPU FP64 cross-reference discrepancy is $1.95\times10^{-11}$. Selected $n\leq16$ first-vector diagnostics report normalized fidelity, norm and particle-number moments. Schmidt entropy/effective rank are sampled only at selected cuts and layer boundaries; they do not certify every intermediate bond.

For two-layer H4, capped MPS reaches the physical maximum bond 16 but still has a worst error of $2.89\times10^{-5}$ at the fixed cutoff. Full LiH with cap 64 reaches $3.30\times10^{-6}$. Larger controlled-HEA MPS cases without dense counterparts remain unverified. Selected benchmark CPU FP32 chemistry states have normalized fidelity errors below $3.82\times10^{-12}$, yet full LiH Aer norm drift reaches $2.54\times10^{-4}$. The separate numerical-readout check in Table~\ref{tab:normalization} obtains the norm from the same gate template and includes its timing cost.

\subsection{Reuse and supplementary controls}
\label{app:controls}
Reuse/batching timings are retained in the deposited workflow tables. For the 28-qubit local-observable control, exact backwards-cone reduction uses eight effective qubits: 0.399\,ms versus 8.782\,s for dense Aer, with $2.52\times10^{-14}$ worst compared difference. This concerns the local output; computational-basis terminal shots likewise do not estimate non-diagonal molecular energies.

\subsection{Validation inputs and recorded timing components}
The molecular readout check uses seeds 72001--72003 and eight vectors per seed for H4 at 1.0\,\AA\ and full LiH, both at two layers. Dense FP64 reference discrepancies across the tested SDKs are at most $2.19\times10^{-12}$. MPS validation uses seeds 73001--73003, eight vectors each, with dense references. Saved task-specific profiling records retain 8/36 scalars per candidate; each deployment records and compares all 3,600. LiH profiling/deployment seeds are 74001/75001; all launches reuse the same held-out stream to measure launch variation. Startup includes parameter generation and hashing; cold means new worker processes, without flushing OS caches. Median last/first full 20-call chunk-time ratios across workers and launches are 0.999/1.003/0.992 for NumPy/Aer/Lightning. The three deployments repeat 3,600 unique held-out vectors, rather than providing 10,800 independent vectors. Task-specific profiling costs sum startup and parent-observed rounds, excluding teardown and gaps; deployment RSS sums sampled child processes and excludes the supervisor.

Table~\ref{tab:components} reports the available NumPy timing decomposition. LiH's native NumPy path spends 251.6\,ms in evolution and 51.5\,ms in its observable, whereas the 18-qubit HEA observable costs 1.0\,ms. These are matched recorded CPU calls. Lightning and CUDA-Q expose no corresponding phase split in the archive, so this decomposition cannot assign the CPU/GPU reversal to a particular phase.
\begin{table}[t]
\centering\footnotesize\setlength{\tabcolsep}{3.5pt}
\caption{NumPy FP64 call components on Barbora, one thread, in milliseconds. Mean of per-seed mean matched last-vector calls across timing rounds. Other includes Python adapter/batch and timer overhead. These are CPU component measurements, not GPU or Lightning kernel profiles.}
\label{tab:components}\begin{tabular}{lrrr}
\toprule
Workload & Evolution & Observable & Other \\
\midrule
LiH, 12q, $p=4$ & 251.61 & 51.54 & 0.020 \\
TFIM block, 20q, $p=4$ & 1285.62 & 665.45 & 0.041 \\
HEA, 18q, $L=4$ & 229.17 & 1.00 & 0.022 \\
\bottomrule
\end{tabular}

\end{table}

\subsection{Auxiliary power and energy evidence}
\label{app:power}
An auxiliary sequential check changed GPU~0/GPU~1 limits from 400/400 to 600/400\,W and reused the same three seeds: the pooled paired FP64 ratio $T_{400}/T_{600}=1.0007$ showed no material timing benefit. The main figures use the 600/400\,W execution; the two executions are not treated as six independent seeds. Board-power telemetry and CPU RAPL counters have different scopes, include setup and unequal call counts, and do not establish whole-system energy per evaluation; a power ceiling is not consumed power \cite{nvidiasmi}.

At assumed constant 300\,W, the executed Aer/NumPy/Lightning LiH deployments correspond to 7.67/1.89/1.58\,Wh, excluding task-specific profiling. Task-specific profiling adds another 2.97\,Wh under that assumption. Lightning's runtime advantage over NumPy implies lower deployment electricity only if its mean whole-node power is less than 1.20 times NumPy's. These are conditional runtime conversions; measured whole-system power and site overhead are required for an electricity comparison.

\FloatBarrier
\bibliographystyle{elsarticle-num}
\bibliography{references}
\end{document}